\pdfoutput=1
\documentclass[aps,prl,twocolumn,superscriptaddress,a4paper,floatfix,nobibnotes]{revtex4}
\usepackage{siunitx}
\usepackage{color}
\usepackage{sfmath}
\usepackage{graphicx}
\usepackage{amsmath,amssymb}
\usepackage{hyperref}

\begin{document}
\title{Breakdown of the Wiedemann-Franz law in a unitary Fermi gas}
\author{Dominik Husmann}
\affiliation{Institute for Quantum Electronics, ETH Zurich, 8093 Zurich, Switzerland}
\author{Martin Lebrat}
\affiliation{Institute for Quantum Electronics, ETH Zurich, 8093 Zurich, Switzerland}
\author{Samuel H\"ausler}
\affiliation{Institute for Quantum Electronics, ETH Zurich, 8093 Zurich, Switzerland}
\author{Jean-Philippe Brantut}
\affiliation{Insitute of Physics, EPFL, 1015 Lausanne, Switzerland}
\author{Laura Corman}
\affiliation{Institute for Quantum Electronics, ETH Zurich, 8093 Zurich, Switzerland}
\author{Tilman Esslinger}
\affiliation{Institute for Quantum Electronics, ETH Zurich, 8093 Zurich, Switzerland}

\begin{abstract}
We report on coupled heat and particle transport measurements through a quantum point contact (QPC) connecting two reservoirs of resonantly interacting, finite temperature Fermi gases. After heating one of them, we observe a particle current flowing from cold to hot. We monitor the temperature evolution of the reservoirs and find that the system evolves after an initial response into a non-equilibrium steady state with finite temperature and chemical potential differences across the QPC. In this state any relaxation in the form of heat and particle currents vanishes. From our measurements we extract the transport coefficients of the QPC and deduce a Lorenz number violating the Wiedemann-Franz law by one order of magnitude, a characteristic persisting even for a wide contact. In contrast, the Seebeck coefficient takes a value close to that expected for a non-interacting Fermi gas and shows a smooth decrease as the atom density close to the QPC is increased beyond the superfluid transition. Our work represents a fermionic analog of the fountain effect observed with superfluid helium and poses new challenges for microscopic modeling of the finite temperature dynamics of the unitary Fermi gas.
\end{abstract}

\maketitle

The interplay between heat and matter currents in a many-body system sheds light on its fundamental properties and the character of its excitations. Transport measurements are a particularly important probe in presence of strong interactions and high temperatures $T$, when a microscopic model is absent or computationally intractable. Phenomenologically, the dependence of the currents on external biases is captured by transport coefficients, such as the particle conductance $G$ or the thermal conductance $G_T$. They determine the ability of a system to relax towards equilibrium at long times, and give unique information on its physical nature. For instance, the Wiedemann-Franz law states that the ratio $G_T /T G \equiv L$, the Lorenz number, takes a universal value for all Fermi liquids in the low-temperature limit. Therefore any breakdown signals physics going beyond a Fermi liquid behavior. In addition, measuring the coupling between heat and particle currents, the Seebeck or Peltier effects, gives direct access to the entropy carried by one transported particle and sensitively probes the energy-dependence of the transport processes. Numerous studies have documented the importance of such measurements \cite{he_advances_2017,benenti_fundamental_2017}, both for realizing efficient thermoelectric materials \cite{snyder_complex_2008} and for understanding correlated systems \cite{wakeham_gross_2011,wang_spin_2003}.

A cold atomic Fermi gas in the vicinity of a Feshbach resonance is a fundamental example of a strongly correlated Fermi system. Owing to the control offered by laser manipulation, its trapping potential can be shaped into custom geometries such as a two-terminal configuration, allowing to measure transport coefficients \cite{krinner_two-terminal_2017}. Previous studies of the unitary Fermi gas have charted out its thermodynamic properties \cite{ku_revealing_2012,nascimbene_exploring_2010}. Recently, transport experiments have observed dissipation processes occurring in the presence of a weak link, such as vortex nucleation in a Josephson junction and multiple Andreev reflections \cite{burchianti_connecting_2018,valtolina_josephson_2015,husmann_connecting_2015}, and heat waves in the form of second sound have been observed \cite{sidorenkov_second_2013,hou_first_2013}.
However, the thermoelectric coupling between heat and particle currents in the unitary regime has not been experimentally addressed so far. Such a study is particularly relevant for applications to cooling protocols as well as for singling out the contribution of fermionic particles to heat flow \cite{kikkawa_magnon_2016,kim_heat_2012, grenier_peltier_2014,papoular_increasing_2012}.
Indeed, in contrast to solid state systems, where the lattice melts at high temperatures, the unitary Fermi gas realized in cold atoms remains free of lattice phonons at all temperatures.

In this paper, we report on measurements of heat and particle transport through a quantum point contact connecting two reservoirs of strongly correlated Fermi gases across the superfluid transition. We observe the evolution of an initially imposed temperature imbalance for equal atom numbers in a two-terminal Landauer configuration \cite{krinner_two-terminal_2017}. In general, coupled particle and heat currents tend to pull a system towards thermodynamical equilibrium. However, here the system evolves towards a non-equilibrium steady state (NESS) within the time scale of the experiment. While typically a NESS is associated to stationary states of open systems \cite{labouvie_bistability_2016}, here it can also describe our experiments thanks to the presence of dissipation and thermodynamic driving forces.
Our results sharply contrasts with previous experiments observing heat transport with weakly interacting atoms \cite{brantut_thermoelectric_2013}. There a single time constant was found to describe the dynamics for temperature and particle relaxation.

Here, our observations reveal a strong separation of heat and particle transport timescales, resulting in a Lorenz number much lower than the value expected for a Fermi liquid. The paradigmatic system supporting suppressed heat transport is the so-called superleak in liquid bosonic helium II. Heating one side of the superleak yields the fountain effect, where both the Seebeck coefficient and the thermal conductance vanish. Our observations represent a fermionic analog to the fountain effect, where the QPC takes the role of the superleak with very low thermal and particle conductance in the non-interacting limit. We however measure a finite Seebeck coefficient even in the superfluid regime, calling for a description of the transport process going beyond the standard two-fluid model. 
\begin{figure}[htb]
\centering
\includegraphics[width=.8\linewidth]{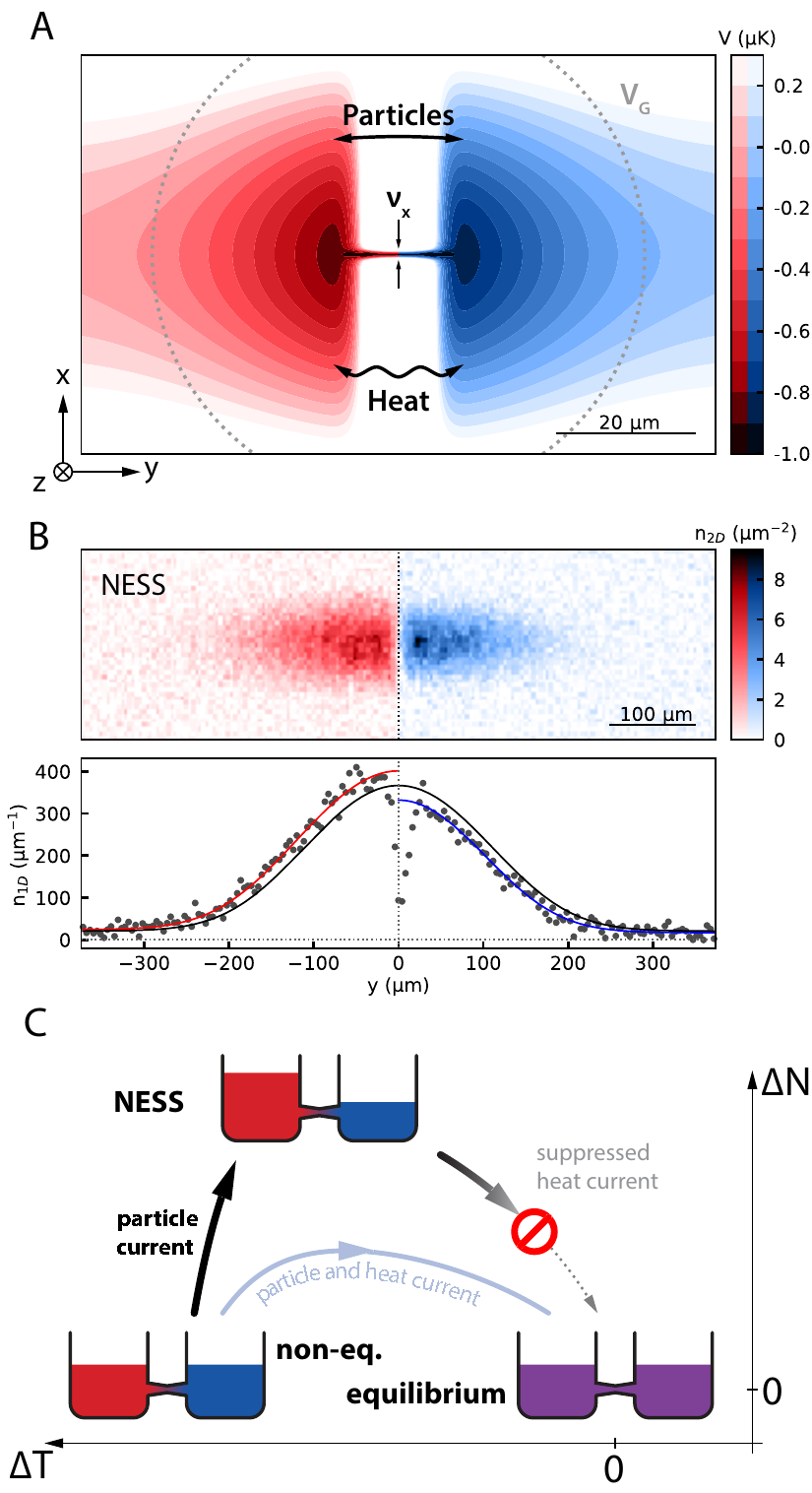}
\caption{
		\textsl{(A)} Particle and heat exchange between two reservoirs containing fermions with resonant interactions, mediated by the QPC which is characterized by the confinement frequencies \(\nu_x\) and \(\nu_z\) (not shown), and a gate potential \(V_G\).
		We indicate equipotential lines at the channel connecting the hot (red) and cold (blue) reservoirs.
		\textsl{(B)} Absorption picture and density profile after 1~ms of time-of-flight. Here, a NESS has been reached after 4~s of transport. Solid lines indicate fits (SI Appendix), from which we deduce both the atom numbers \(N_{\rm R}=35(1)\cdot 10^3\), \(N_{\rm L}=51(1)\cdot 10^3\), and temperatures \(T_{\rm R}=163(9)\)~nK and \(T_{\rm L}=231(8)\)~nK. 
		\textsl{(C)} Schematics of the thermodynamic equilibration process in \( (\Delta T,\Delta N ) \)-space with \(\Delta T\) and \(\Delta N\) increasing along arrow direction. In general, a temperature-biased non-equilibrium state evolves towards thermodynamic equilibrium along a trajectory indicated in light blue. In the particular case of vanishing heat diffusion, the system evolves predominantly along the \(\Delta N\)-axis to a NESS where the absence of thermal relaxation processes prevents evolution to thermodynamic equilibrium.
		}
\label{fig:setup}
\end{figure}

\section*{System}
\label{sec:system}

Our experiment consists of a QPC imprinted onto a cold, unitary Fermi gas of \(^6\)Li atoms, like in our previous work \cite{krinner_observation_2015, husmann_connecting_2015}.
We form the QPC using two far-detuned repulsive laser beams with a line of zero intensity in the center, resulting in a region of tight harmonic confinement with typical frequencies of 
\(\nu_x = 20.2(6)\)\ kHz 
 and 
\(\nu_z = 12.9(5)\)\ kHz 
(see Fig.~\ref{fig:setup}A) setting an energy spacing between transverse modes of \( h\nu_{x(z)}\).
This mesoscopic structure connects two initially decoupled reservoirs, labeled as left (L) and right (R), and enables the transport of particles and heat. 
The reservoirs contain a mixture of the lowest and third lowest hyperfine state, with typically \( N = N_{\rm L} + N_{\rm R} = 97(4)\cdot 10^3 \) 
atoms in each spin state, temperatures of 
\(\bar{T} = (T_{\rm L} + T_{\rm R}) /2= 184(8)\)~nK 
 and chemical potentials of 
\(\bar{\mu} = (\mu_{\rm L} + \mu_{\rm R}) /2 = 272(24)\ \mathrm{nK} \cdot k_{\rm B}\). 
Here \( N_i \), \( T_i \) and \( \mu_i\) with \( i = \mathrm{L,R}\) indicate the atom number, temperature and chemical potential of the individual reservoirs. We control the density inside and close to the QPC by varying the power of an additional laser beam which creates an attractive gate potential \(V_G\). 
At this temperature and chemical potential, up to a few transverse modes in the \(x\)- and \(z\)-direction are populated (SI Appendix).

We bring the system out of equilibrium by heating either of the reservoirs using an intensity-modulated laser beam focused on the reservoir, while maintaining the QPC closed. This results in a temperature difference up to 
\( \Delta T = T_{\rm L} - T_{\rm R} = \pm 83 \)~nK. 
In each of the reservoirs, seen as half-harmonic traps, atom number \(N\) and internal energy \( E \) are obtained from density profiles (see Fig.~\ref{fig:setup}B).
These quantities can be converted to any other thermodynamic variable such as temperature \(T\) or chemical potential \( \mu \) using the unitary equation of state (EoS). Here the reduced chemical potential 
\( q = \bar{\mu} / k_{\rm B} \bar{T} \approx 1.5 \) 
in the reservoirs is below the superfluid transition point at \( q_c = 2.5\) \cite{hou_first_2013}.
By tuning the gate potential \(V_G\) the degeneracy can be increased in the vicinity of the QPC. The gate beam thus acts as a local dimple creating superfluid regions close to the QPC (SI Appendix). We perform a transport experiment by opening the QPC for a variable time \(t\) between 0~s and 4~s, and subsequently measuring \(\Delta N(t)\) and \(\Delta T(t)\). For convenience, we omit to write explicitly the time-dependence of these quantities. Particle and heat exchange between the reservoirs is enabled during this time, leading to relaxation dynamics depicted in Fig.~\ref{fig:setup}C.

\section*{Dynamics}
\begin{figure}[htb]
	\centering
		\includegraphics[width=.8\linewidth]{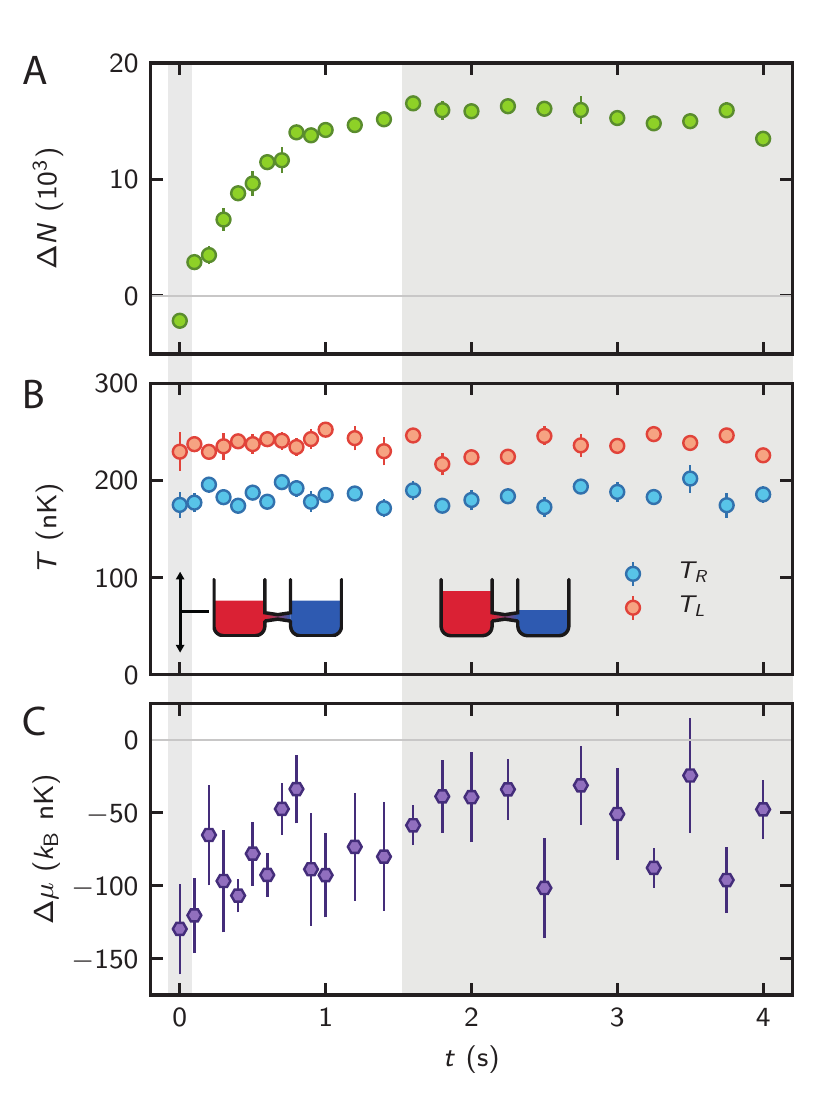}
		\caption{
		\textsl{(A)} Evolution of particle number imbalance \(\Delta N\), \textsl{(B)} temperatures in the left (red) and right (blue) reservoir \( T_{\mathrm{L,R}}\)  and \textsl{(C)} chemical potential bias \(\Delta \mu\) as a function of time \(t\) for an initial temperature imbalance of \( \Delta T_0=49(8) \)~nK. Within the first 1.5~s a particle imbalance builds up, while the temperature bias shows no measurable evolution. On the same timescale, \(\Delta \mu\) decreases to a finite non-zero value, leading to a NESS with finite \(\Delta N\) and \(\Delta \mu\). In this configuration the high-density regions close to the QPC are superfluid with local reduced chemical potentials of \(q_{\rm L}=3.2(4)\) on the left and \(q_{\rm R}=5.0(5)\) on the right side.
	}
	\label{fig:transients}
\end{figure}

Figure~\ref{fig:transients} presents a typical time evolution of \(\Delta N\), \( T_{\rm R, L}\), and \( \Delta \mu\) during transport. The gate potential and transverse frequencies were set to \( V_G = 1.00(3) \)~\(\mu\)K\(\cdot k_\mathrm{B}\) and \(\nu_x = 20.2(6)\)~kHz respectively. The high-density regions close to the QPC are superfluid in this configuration.
The initial state was prepared with \(\Delta T_0 = \Delta T (t=0) = 49(8)\)~nK and  \(\Delta N_0 = 0\), with the equilibrium thermodynamics of the reservoirs yielding \(\Delta\mu_0 = \mu_{\rm L}-\mu_{\rm R} \approx -130(31)\ \mathrm{nK} \cdot k_{\rm B}\).
During the first \(1.5\)\ s, a relative particle imbalance \(\Delta N / N \approx 0.16(1)\) rapidly builds up, leading to a decrease of \(|\Delta\mu|\) while \(\Delta T\) is constant within experimental resolution. This evolution is driven by the finite value of  \(\Delta\mu\), resulting in a large current from the cold to the hot reservoir. It dominates over the much weaker thermoelectric current from the hot to the cold reservoir, originating from the energy dependence of the transmission through the QPC. This contrasts with previous observations with weakly interacting atoms  \cite{brantut_thermoelectric_2013, grenier_thermoelectric_2016} where the thermoelectric current was the dominating contribution.

After a typical timescale \(\tau_+\), \(\Delta T\) and \(\Delta N\) reach a steady state, strongly departing from thermodynamical equilibrium. Interestingly, the decline in \(\Delta \mu\) stops at a non-zero value \(\Delta\mu_s \approx -55(8)\ \mathrm{nK} \cdot k_{\rm B}\), which is estimated by taking the average over the data points for times \(t\geq1.6\)\,s. To account for the very weak decrease of  \(\Delta N\) in the second half of the observation time, we introduce a much longer timescale \(\tau_-\) describing the decay of \(\Delta N\) and \(\Delta T\) back to zero. This timescale corresponds to the thermal equilibration process shown in Fig.~\ref{fig:setup}C, and our observation shows that \(\tau_- \gg \tau_+\).

To provide a quantitative understanding of the time evolution of the system, we use a phenomenological model based on linear response. While such an approach is known to fail in the lowest temperature regimes, where nonlinear current-bias relations have been observed \cite{husmann_connecting_2015}, we find that it describes our observations well (SI Appendix), and allows for comparison between different QPC parameters. 
In this framework, the particle current $I_N = -1/2 \cdot {\rm d}\Delta N/{\rm d}t$ and entropy current $I_S = -1/2 \cdot {\rm d}\Delta S/{\rm d}t$ are expressed as a function of the differences in chemical potential \(\Delta \mu\) and temperature \(\Delta T\) between the reservoirs \cite{grenier_thermoelectric_2016,goupil_thermodynamics_2011}:

\begin{equation}
    \begin{pmatrix}
    I_N  \\
    I_S  
    \end{pmatrix}
    = G
    \begin{pmatrix}
    1                   &   \alpha_{\rm{c}} \\
    \alpha_{\rm{c}}    & L + \alpha_{\rm{c}}^2
    \end{pmatrix}
    \cdot
    \begin{pmatrix}
    \Delta\mu\\
    \Delta T 
    \end{pmatrix}
    \label{eq:2by2}
\end{equation}

The transport properties of the channel are captured by its particle and thermal conductances $G$ and $G_T$, which can be combined into the Lorenz number \(L = G_T/(\bar{T} G)\), and its Seebeck coefficient $\alpha_{\rm{c}}$ describing the coupling between particle and entropy currents. 

The absence of the relaxation of temperature and particle imbalance shown in Fig.~2 implies a very low heat conductance. According to the first law of thermodynamics the energy flow \( I_E \) can be expressed as:
\begin{equation}
    I_E = \bar{T} \cdot I_S + \bar{\mu} \cdot I_N = (\bar{\mu} + \alpha_{\rm c} \bar{T}) I_N + G_T \Delta T ,
\label{eq:J_E}
\end{equation}
where the first term on the right represents work flow and the second term heat flow. Work is associated with the reversible transfer of an average energy per particle \(\bar{\mu} + \alpha_\mathrm{c} \bar{T}\), while irreversible, diffusive heat transfer is proportional to \(\Delta T\), obeying Fick's law. From Fig.~2A, we find \(I_N = 0\) and \( \Delta T > 0 \) for longer times. A direct measurement of \(I_E\) yields a low value for the heat conductance of 
\(G_T=0.2\cdot G_{T,\mathrm{NIF}}\),
 where \( G_{T,\mathrm{NIF}} \) is the conductance expected for a non interacting Fermi gas with the same chemical potential, temperature and gate potential (SI Appendix).

\begin{figure*}
	\centering
		\includegraphics{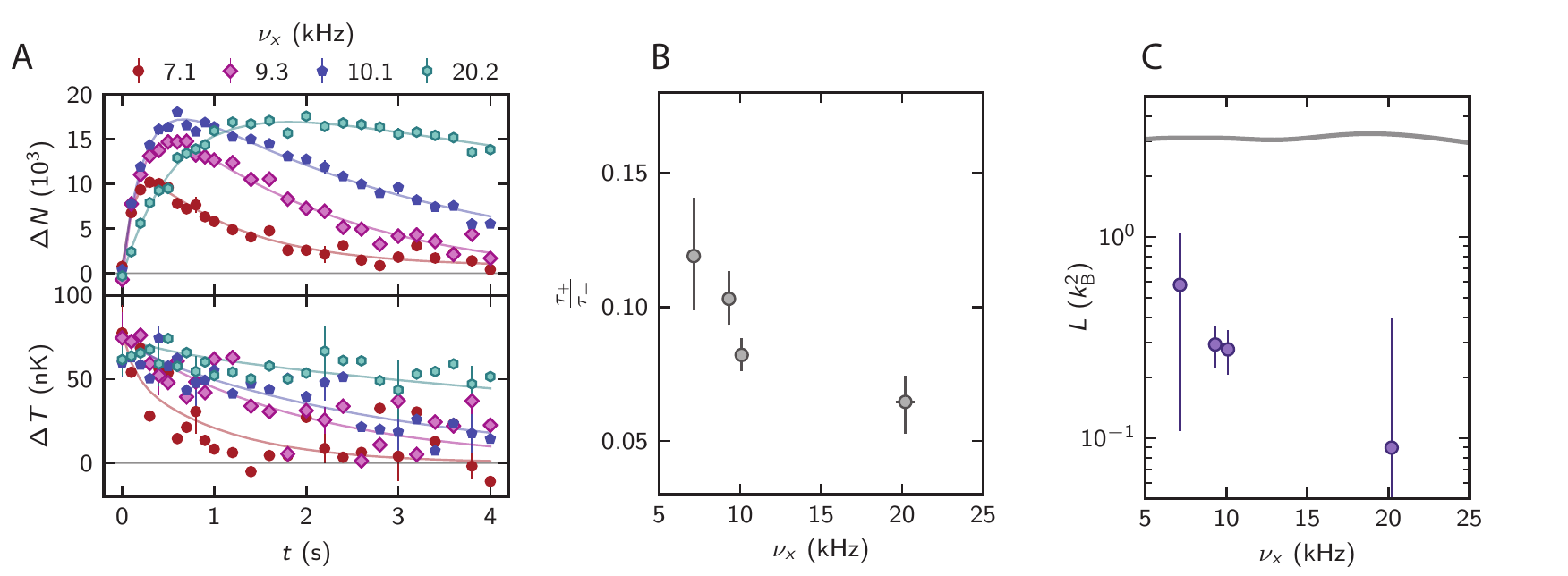}
	\caption{Variation of the transverse confinement frequencies \(\nu_x\).
	\textsl{(A)} Evolution of \(\Delta N\) and \(\Delta T\) for confinements \(\nu_x=7.1(2),\ 9.3(3),\ 10.1(3),\) and \(\ 20.2(6)\ \mathrm{kHz}\) for fixed \(V_G = 1.00(3)\ \mu\mathrm{K}\cdot k_\mathrm{B}\). For high confinement, the system evolves towards a NESS. A decrease in \(\nu_x\) results in increased relaxation to thermodynamical equilibrium. Solid lines represent bi-exponential fits to the data (see \eqref{eq:fits}). 
	The reduced chemical potentials in the left and right reservoir at \(t=0\)~s are \(q_{\mathrm L}=3.4(6) \) and \(q_{\mathrm R}=5.9(7) \), where the values are averaged over the different confinements. Error bars indicate standard errors for every fourth data point. 
	\textsl{(B)} Ratio of the timescales \(\tau_+,\ \tau_-\) from bi-exponential fits for various values of the confinement. While both timescales increase with confinement, their ratio decreases, indicating a strong change in the transport coefficients for heat and particle transport.
	\textsl{(C)} Experimentally determined Lorenz number \(L\) (violet circles) and Lorenz number expected for an equivalent non-interacting system (gray line) at equal chemical potential and temperature. The measured values lie consistently below the \(\pi^2 / 3\cdot k_{\rm B}^2\) predicted by Wiedemann-Franz law.}
	\label{fig:lorenz_wire}
\end{figure*}

\section*{Transport coefficients}
The transport parameters in mesoscopic system strongly depend on the channel geometry of the channel \cite{imry_introduction_2002}. We investigate this dependency by measuring the dynamics of atom number difference and temperature difference as the channel confinement is reduced, departing from the single-mode regime.
Fig.~\ref{fig:lorenz_wire}A presents the results for four different transverse confinements \(\nu_x\).
For weaker confinements, \(\Delta N\) and \(\Delta T\) equilibrate to zero for long times. This is expected as the geometric contact between the two reservoirs increases in size, leading to higher particle and diffusive heat currents. 

We fit the time traces with the solutions from the linear response model in \eqref{eq:2by2}, which are bi-exponential functions where we fixed \(\Delta N(t=0)=0\) and \(\Delta T(t=0)=\Delta T_0\) according to our preparation:
\begin{align}
    \Delta N(t) &= A[\exp(-t/\tau_+) - \exp(-t/\tau_-)] \\
    \Delta T(t) &= B\exp(-t/\tau_+) + (\Delta T_0 - B)\exp(-t/\tau_-).
		\label{eq:fits}
\end{align}
The fit parameters \(\tau_+,\ \tau_-,\ A\) and \(B\) are functions of the transport coefficients of the channel \(\alpha_{\rm c},\ G,\ L\) and the thermodynamics of the reservoirs through their compressibility, heat capacity and dilatation coefficient. The fit is performed simultaneously on both \(\Delta N\) and \(\Delta T\), normalized with the statistical uncertainty of the data. We find two timescales \(\tau_{+}\) and \( \tau_{-}\) that differ by one order of magnitude, a feature that remains even for fast equilibration at weak confinement (see Fig.~\ref{fig:lorenz_wire}B). 
Consequently, each timescale can be mapped to the relaxation dynamics of heat (\( \tau_- \)) and particles (\( \tau_+ \)) (SI Appendix). Within this linear response solution, the direction and magnitude of the currents result from a competition between the transport properties of the channel and the thermodynamic response of the reservoirs (SI Appendix), a feature that was encountered already for the weakly interacting Fermi gas in \cite{brantut_thermoelectric_2013}. 

\subsection*{Lorenz number}
The relative weight of particle and heat conductance is captured by the Lorenz number \( L \). Direct conversion of the fit parameters in \eqref{eq:fits} to \( L\) is however not possible because the bi-exponential model is ill-conditioned.
Instead we express \(L\) by estimating \(G\) and \(G_T\) from particle and energy currents \(I_N\) and \(I_E\) obtained from the data. The thermal conductance and Seebeck coefficient are given by \(G_T = I_E /\Delta T \) and \(\alpha_{\mathrm{c}} = -\Delta\mu / \Delta T \) at the point of vanishing particle current (see \eqref{eq:2by2}). The conductance \(G\) is calculated for short transport times where we obtain \(G = I_N / (\Delta\mu + \alpha_{\rm c}\Delta T)\) (see SI Appendix for details).

The estimates of the Lorenz number are presented in Fig.~\ref{fig:lorenz_wire}C, together with the expected value for a non-interacting QPC obtained through Landauer theory with equivalent chemical potential, temperature and channel properties \cite{grenier_thermoelectric_2016}. For all values of \(\nu_x\), \(L\) is much smaller than in the non-interacting case, which approaches the Wiedemann-Franz law \(L_{\rm WF}=\pi^2 / 3\cdot k_{\rm B}^2\). Our observations thus violate the Wiedemann-Franz law by an order of magnitude. This law roots in having the same carriers for charge and heat, and is robust to moderate interactions, where the system can be described by a Fermi liquid. Deviations from this law may appear when Fermi liquid theory breaks down, as encountered for example in strongly correlated 1D systems, where the Lorenz number can either increase \cite{wakeham_gross_2011} or decrease \cite{lee_anomalously_2017,bruin_similarity_2013,hartnoll_theory_2015}. This is in line with our previous work, which showed that the conductance of the strongly interacting Fermi gas close to the critical point strongly differ from the predictions of the Landauer formula \cite{krinner_mapping_2016,husmann_connecting_2015}.

\begin{figure*}[htb]
	\centering
		\includegraphics{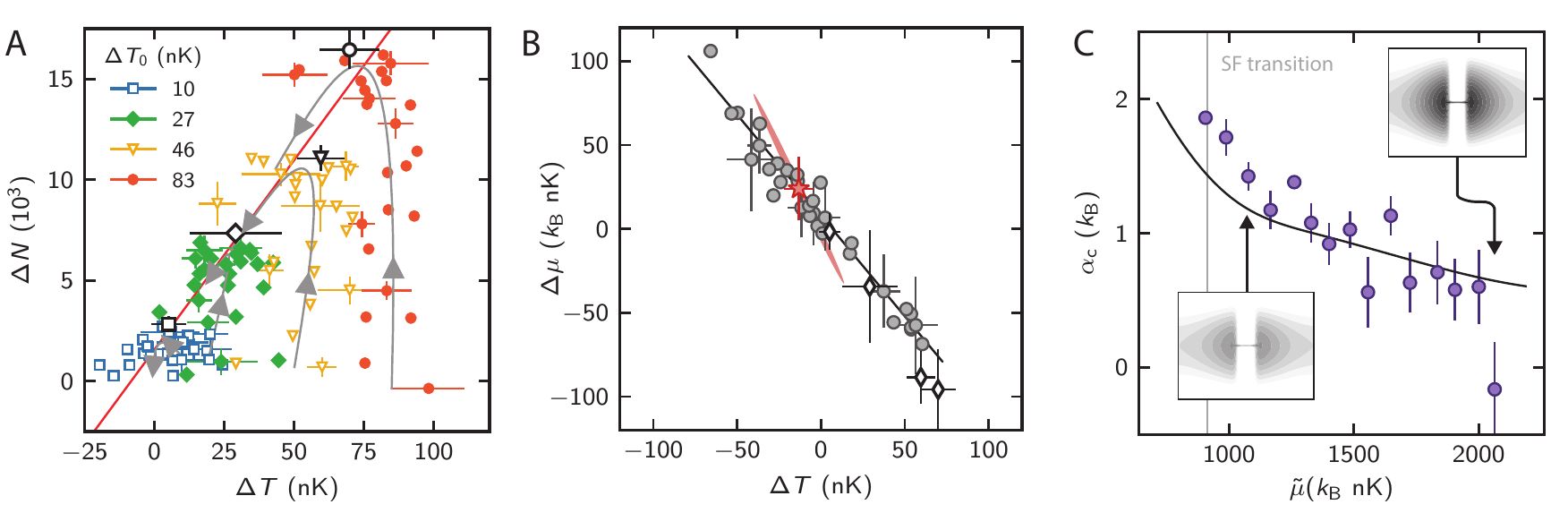}
	\caption{
	\textsl{(A)}	Transients in \( ( \Delta T ,\Delta N ) \)-space for various values of initial heating \(\Delta T_0 = 10(5),\ 27(7),\ 46(8)\) and \(83(7)\)~nK at fixed \( \nu_x = 20.2(6)\)~kHz and \(V_G = 1.00(3)\ \mathrm{nK}\cdot k_\mathrm{B}\). Gray lines are fits with Eq.~\eqref{eq:fits} for \(t=0-8\)~s, where the arrows indicating the time progress. The NESS is at different stopping points (black) depending on the initial heating. Error bars indicate standard errors for every fourth data point.
	\textsl{(B)} Detailed study of \(\Delta \mu\) vs. \(\Delta T \) (gray circles) at \(t=2\)~s, where particle current is zero. The black line indicates a linear fit to the data, with the slope representing the Seebeck coefficient \( \mathrm{d}\Delta\mu / \mathrm{d}\Delta T = -\alpha\). Included in the fits is the covariance of the data points, shown by the red shaded region for a selected data point (red star) representative for the full data set (for details on the fit see SI Appendix). The four stopping points from (A) are indicated as black diamonds. Error bars indicate standard errors for every fourth data point.
	\textsl{(C)} Measurement of the Seebeck coefficient for various values of the gate potential \(V_G\). The black line shows the prediction for an equivalent non-interacting system (equal temperature and chemical potential). The transition point in the pockets from normal to superfluid is indicated as a gray line.}
	\label{fig:seebeck_dimple}
\end{figure*}

\subsection*{Seebeck coefficient}
The steady state observed in Figs.~\ref{fig:transients} and \ref{fig:lorenz_wire}A allows to relate chemical potential and temperature differences to the Seebeck coefficient.
From the stationary state realized in Fig.~\ref{fig:transients}, we find $\alpha_\mathrm{c} =-\Delta\mu / \Delta T =  1.1(2) k_\mathrm{B}$. Here we rely on a linear relation between the stationary value of \(\Delta T\) and \(\Delta \mu\) (see \eqref{eq:2by2}) at the times where particle current \(I_N = 0\) vanishes (gray area in Fig.~\ref{fig:transients}). We measured \((\Delta N,\Delta T)\) for different values of the heating (see Fig.~\ref{fig:seebeck_dimple}A for time traces), and convert them to \((\Delta \mu, \Delta T)\).

For the specific case of \( V_G = 1.00(3)\ \mu\mathrm{K}\cdot k_\mathrm{B}\) and \(\nu_x = 20.2(6)\)~kHz, we characterize the quasi-steady state for a transport time of \(t=2\) s, and find a linear relation between \(\Delta \mu\) and \(\Delta T\) (see Fig.~\ref{fig:seebeck_dimple}B). This confirms that the linear model \eqref{eq:2by2} constitutes an adequate description of our system. The linear relation yields a Seebeck coefficient \(\alpha_\mathrm{c}= 1.2(2) k_{\rm B} \). To check that the measurement of \( \alpha_\mathrm{c} \) does not depend on the precise value of the transport time, we repeat the measurement at \( t = 4\)~s and find consistent values.
This value is very close to the  case of a one-dimensional quantum wire in the non-interacting regime, where one expects \(\alpha_{\rm NIF}=1 k_{\rm B}\) (SI Appendix).
We further investigate the Seebeck coefficient by increasing the attractive gate potential \(V_G\) centered on the QPC. This method has two consequences on the two-terminal system:
(i) it probes the single-particle energy dependence of the transport parameters by increasing the number of available modes in the QPC
(ii) the density in the vicinity of the QPC is modified by tuning the chemical potential, locally increasing the superfluid gap.
We measure \(\Delta\mu\) and \(\Delta T\) for various heating strengths when \(I_N = 0\) as for Fig.~\ref{fig:seebeck_dimple}B, and deduce \(\alpha_\mathrm{c}\).

Figure~\ref{fig:seebeck_dimple}C shows \(\alpha_\mathrm{c}\) as a function of the chemical potential modified by the attractive gate \(\tilde{\mu} = \bar{\mu} + V_G\). The Seebeck coefficient decreases from a value slightly below \(2 k_{\rm B} \) at \( \tilde{\mu} =  0.91(3)\ \mu\mathrm{K}\cdot k_{\rm B}\)\ to a value close to zero for \(\tilde{\mu}>2.06(6)\ \mu\mathrm{K}\cdot k_{\rm B}\).
A similar decrease of \( \alpha_\mathrm{c} \) is theoretically expected for a non-interacting QPC (see black curve in Fig.~\ref{fig:seebeck_dimple}C), and is explained there by an increase of the number of 1D channels available for the transport of single particles. This similarity is surprising as transport coefficients in this regime close to the superfluid transition have shown order of magnitude deviations from the Landauer model \cite{krinner_mapping_2016,husmann_connecting_2015}.
The residual deviation from the non-interacting curve --- manifested in the faster decrease with \(\tilde{\mu}\) --- is compatible with the expectations for a BCS superfluid close to the transition point, where a smooth decrease of \(\alpha_\mathrm{c}\) from the non-interacting value to zero is expected \cite{guttman_thermoelectric_1997}.  

\section*{Discussion}
The coexistence of a vanishing Lorenz number and a finite Seebeck coefficient leading to a NESS at finite \(\Delta \mu\) distinguishes our observations from the fountain effect seen in superfluid helium II \cite{allen_flow_1938,kapitza_viscosity_1938}. 
There, two vessels are connected by a macroscopic duct, called superleak. Heating one of them induces a current from cold to hot until a steady state with different temperatures and pressures is reached.
In a two-fluid description of the superleak, viscosity prevents the normal, entropy-carrying fraction of the fluid to cross while allowing the entropy-less, superfluid fraction to flow and equilibrate the chemical potentials, \(\Delta \mu = 0\) \cite{karpiuk_superfluid_2012,sekera_thermoelectricity_2016}. On the contrary, our system is characterized by a non-zero Seebeck coefficient \(\alpha_\text{c} = I_S /I_N \) in the limit \(L = 0\), indicating a mean entropy transported by each particle of about \(1 k_\mathrm{B}\).
In addition, our QPC is ballistic and not diffusive, and both its high resistance for a normal Fermi gas \cite{krinner_observation_2015} and its low resistance for a superfluid \cite{husmann_connecting_2015} is predominantly determined by quantum effects. The hydrodynamic models describing the fountain effect are therefore expected to break down.  Although thermoelectric transport across a Josephson junction is well understood within BCS theory \cite{guttman_thermoelectric_1997,giazotto_josephson_2012}, and interaction effects for bosonic and fermionic low-dimensional systems have been studied \cite{filippone_violation_2016,rancon_bosonic_2014}, no model has so far been proposed to describe the unitary Fermi gas at a QPC.

The non-vanishing \(\alpha_{\mathrm{c}}\) associated with a low Lorenz number suggests that our QPC considered as a thermoelectric device has a high efficiency.
Within this framework, the time evolution of the reservoirs describes an open thermodynamic cycle. There, the system acts first as a thermoelectric cooler, where a chemical potential difference drives convection heat from the cold to the hot reservoir, followed by a thermoelectric engine part, where the temperature difference drives particles from a lower to a higher chemical potential and hence produces work.
As the transverse confinement frequency \( \nu_x \) is increased, both processes slow down and evolution gets closer to reversibility, resulting in a decrease of the output power \(P=I_N \Delta \mu\) and in a better conversion efficiency between work and heat.
This efficiency is determined by a dimensionless figure of merit \( ZT = \alpha_{\rm c}^2/ L \) \cite{goupil_thermodynamics_2011} which is on the order of 14(8) for the largest confinement \( \nu_x = 20.2(6) \) kHz, where the large errors stem from uncertainty in \(\Delta T\) and \(\Delta \mu\). Currently the best thermoelectric materials have figures of merit on the order of 3 to 5 \cite{snyder_complex_2008,tan_non_equilibrium_2016}.
Further considerations on the efficiency are given by the comparison to the Carnot efficiency (SI Appendix).

Our fountain effect setting with fermions provides a conceptual link between the thermoelectric transport witnessed in electronic devices and the bosonic fountain effect observed with helium II. Its anomalous features --- exceptionally small Lorenz number and finite Seebeck coefficient --- shed new light on the out-of-equilibrium properties of the unitary Fermi gas, but also underline the necessity of a better understanding of strongly correlated systems at finite temperatures. These results portend potential applications to ultracold atoms, such as the realization of novel cooling schemes.

\begin{acknowledgments}
We thank H.~Aoki, A.~Georges, T.~Giamarchi, L.~Glazman, D.~Papoular, S.~Pershoguba, S.~Uchino and W.~Zwerger for discussions;
B.~Frank for providing the data on the superfluid gap;
and B.~Braem and P.~Fabritius for careful reading of the manuscript.
We acknowledge financing from Swiss NSF under division II (Project Number 200020\_169320 and NCCR-QSIT), Swiss State Secretary for Education, Research and Innovation Contract No. 15.0019 (QUIC), ERC advanced grant TransQ (Project Number 742579) and ARO-MURI Non-equilibrium Many-body Dynamics grant (W911NF-14-1-0003) for funding. J.-P.B. is supported by the ERC starting grant DECCA (Project Number 714309) and the Sandoz Family Foundation-Monique de Meuron program for Academic Promotion. L.C. is supported by ETH Zurich Postdoctoral Fellowship and Marie Curie Actions for People COFUND program.
\end{acknowledgments}

\section*{Materials and Methods}
\subsection*{Preparing the cloud and QPC}
We prepare an elongated cloud of fermionic \( ^6 \)Li atoms in a balanced mixture of the lowest and third lowest hyperfine state in a hybrid configuration of a far-detuned 1064~nm dipole trap and a harmonic magnetic trap, confining the atoms along the transverse (\(x,\ z\)) and longitudinal (\(y\)) direction respectively.
We evaporatively cool down the cloud by reducing the trap depth from
\SI{6}{\micro\kelvin} to
\SI{3}{\micro\kelvin} 
on a broad Feshbach resonance at 689~G. After a final tilt evaporation step \cite{hung_accelerating_2008} along the \(z\)-direction, the cloud reaches final temperatures of around \SI{184(8)}{\nano\K}.
The trap frequencies during transport are \( \nu_{\mathrm{r}x} = \SI{318.5}{\hertz} \), \( \nu_{\mathrm{r}y} = \SI{28.4}{\hertz} \) and \( \nu_{\mathrm{r}z} = \SI{255.9}{\hertz} \).
A repulsive lightsheet beam at 532~nm created by a \(\pi\)-phase plate confines the cloud in the center in \(z\)-direction with a longitudinal \(1/e^2\)-waist of
\(w_{\mathrm{LS},y} = \)\SI{30(1)}{\um}. 
An orthogonal 532~nm beam of waist \(w_x = \)\SI{5.49(1)}{\um} in a split-gate shape with an intensity node in the center, realized with a transmission mask imaged onto the atom plane, confines the atoms in the \(x\)-direction. The two transverse confinements effectively lead to a quasi-1D constriction with trapping frequencies \(\nu_x=20.2(6)\)~kHz and \(\nu_z = 12.9(5)\)~kHz.

\subsection*{Transport}
An amplitude-modulated beam at a wavelength of 767~nm is directed on one of the reservoirs, parametrically heating it up. The modulation frequency is optimized experimentally to \(\nu_\mathrm{mod}=125\)~Hz, which is on the order of the transverse trapping frequencies \( \nu_{\mathrm{r}x} \) and \( \nu_{\mathrm{r}z} \) of the dipole trap. The position of this beam is controlled by a piezo-steered mirror and can be shifted to either reservoir. The same beam is centered on the QPC during transport and acts there as an attractive gate potential \(V_G\), locally tuning the density. 
When preparing the reservoirs, transport between the reservoirs is blocked by a repulsive wall beam focused onto the channel. The beam is removed for a variable transport time \(t\) during which exchange between the reservoirs through the channel is enabled. 
After time \( t \) we separate the reservoirs with the wall beam and take absorption images after a short time of flight of 1~ms in the transverse directions. This reduces the densities and allows us to image in the low saturation regime.

\subsection*{Thermodynamic properites of the reservoirs}
From the density profiles we deduce the atom number in each reservoir as well as their internal energy
\begin{equation}
	E = 3 m (2\pi\nu_{\mathrm{r}y}) ^2 \left< y^2 \right>
\label{eq:innerE}
\end{equation}
via the virial theorem for a harmonic trap at unitarity \cite{thomas_virial_2005} and the second moment 
\begin{equation}
	\left< y^2 \right>  = \frac{\int_{-\infty}^{\infty}\mathrm{d}y n_{\mathrm{1D}}(y)y^2}{\int_{-\infty}^{\infty} \mathrm{d}y n_{\mathrm{1D}}(y)}.
\label{eq:2ndmoment}
\end{equation}
of the fitted density distribution \( n_{1D} \) along the longitudinal direction \cite{guajardo_higher-nodal_2013} shown in Fig.~1B. 

Along with the known equation of state of the unitary Fermi gas, these two quantities define all thermodynamic parameters of the individual reservoirs, including their temperatures \(T_{\mathrm{L,R}}\) used in Fig.~\ref{fig:transients}-\ref{fig:seebeck_dimple}. The equation of state is based on measurements in \cite{ku_revealing_2012} and continued towards the normal and degenerate regimes in \cite{hou_first_2013} for a homogeneous gas. We apply local density approximation to obtain the trap-averaged quantities assuming a harmonic potential (SI Appendix).




\clearpage
\newpage 

\makeatletter
\setcounter{section}{0}
\setcounter{subsection}{0}
\setcounter{figure}{0}
\setcounter{equation}{0}
\renewcommand{\theequation}{S\arabic{equation}}
\renewcommand{\thefigure}{S\arabic{figure}}
\renewcommand{\bibnumfmt}[1]{[S#1]}
\renewcommand{\citenumfont}[1]{S#1}

\section*{SI Appendix}

\section{Thermodynamical properties of the reservoirs}
As a result of the scale invariance of the system, the equation of state (EoS) for the density \( n \) of the homogeneous unitary Fermi gas is solely a function of the de Broglie wavelength and the chemical potential normalized by temperature \(q=\mu / k_\mathrm{B}T = \beta\mu\):
\begin{equation}
	n = \frac{f_n(q)}{\lambda^3}.
\label{eq:EoSn}
\end{equation}
The de Broglie wavelength is given by
\begin{equation}
	\lambda = \sqrt{\frac{2\pi\hbar^2}{mk_\mathrm{B} T}},
\label{eq:deBroglie}
\end{equation}
and the thermodynamic density function is
\begin{equation}
	f_n(q)=
	\begin{cases}
		\sum_{i}^{j}b_j j e^{iq},\ q<-0.9 \\
		F_n(q)\left(-\mathrm{Li}_{3/2}(-e^{q})\right),\ -0.9<q<3.5\\
		\frac{(4\pi)^{3/2}}{6\pi^2}\left[\left(\frac{q}{\xi}\right)^{3/2} -\frac{\pi^4}{480}\cdot\left(\frac{3}{q}\right)^{5/2} \right],\ 3.5<q.
	\end{cases}
	\label{eq:EoS}
\end{equation}
The low degeneracy regime is described by a virial expansion in \(\exp\left(q\right)\) with the first four virial coefficients \(b_1=1\), \(b_2=3\sqrt{2} / 8\), \(b_3=-0.291\) and \(b_4=0.065\) \cite{liu_virial_2009_sup,ku_revealing_2012_sup}.
For high degeneracies phonon excitations dominate the excitation spectrum \cite{hou_first_2013_sup}. The intermediate regime interpolating between the two limiting cases has been experimentally measured \cite{ku_revealing_2012_sup} and is contained in the function \( F_n(q) \).

In the presence of a trapping potential \( V(\vec{r}=(x,y,z)) \), the relation holds locally within the local density approximation if one considers the effective dimensionless quantity \(q(\vec{r})=q_0 - \beta V(\vec{r}) \), where \( q_0 \) is \( q \) evaluated at the trap center.

The contributions to the trapping potential relevant for the estimation of the thermodynamical quantities of the reservoirs come from the dipole trap \( V_{\rm DT}\), the magnetic trap \(V_{\rm M}\) and the confinement from the lightsheet \( V_{\rm LS}\):
\begin{equation}
	V(\vec{r}) = V_\mathrm{DT}(\vec{r}) + V_\mathrm{M}(\vec{r}) + V_\mathrm{LS}(\vec{r}).
	\label{eq:trap}
\end{equation}
The individual parts can be approximated as \cite{smith_quasi-2d_2005_sup}
\begin{align}
	V_{\rm DT}(\vec{r}) 	= & V_\mathrm{DT}\left[ 1 - \exp\left(-\frac{2x^2}{w_{x}^2}\right)  
													\exp\left(-\frac{2z^2}{w_{z}^2} \right)\right]
	\label{eq:Vparts1}	\\	
	V_{\rm M}(\vec{r})		= & \frac{1}{2} m \omega_{y}^2 y^2 
	\label{eq:Vparts2}	\\
	V_{\rm LS}(\vec{r})		= & V_\mathrm{LS} \exp\left(-\frac{2y^2}{w_{\mathrm{LS},y}^2}\right) 
	\label{eq:Vparts3}	\\
											&	 \times \exp\left(-\frac{2z^2}{w_{\mathrm{LS},z}^2}\right)\mathrm{erfi}\left(\frac{z}{w_{\mathrm{LS},z}}\right)^2  \nonumber
\end{align}
with the \(1/e^2\)-waists of the dipole trap \(w_x,\ w_z\) and the lightsheet \(w_{\mathrm{LS},y},w_{\mathrm{LS},z}\), and the magnetic confinement trapping frequency \(\omega_y\). 

The total atom number \( N \) of the trapped cloud is obtained by integrating Eq.~\eqref{eq:EoSn} over the full trap potential
\begin{equation}
	N	=	\frac{1}{\lambda^3}\int_{\mathbb{R}^3} f_n\left(q(\vec{r})\right)\mathrm{d}\vec{r} .
	\label{eq:N}
\end{equation}
To compute the chemical potential in each reservoir, we numerically invert Eq.~\eqref{eq:N} and obtain \( \mu = q_0(N,T)\cdot k_\mathrm{B} T \) from the knowledge of the measured quantities \( T \) and \( N \).

For a harmonic trap with mean trap frequency \( \bar{\nu}_\mathrm{r} \), the integral in Eq.~\eqref{eq:N} can be expressed as
\begin{equation}
		N=	\frac{4}{\sqrt{\pi}}\left(\frac{k_B T}{2\pi\hbar\bar{\nu}_\mathrm{r}}\right)^3 N_2(q_0)
\label{eq:N_harm}
\end{equation}
with the dimensionless moments
\begin{equation}
		N_l(q_0) = \int_{0}^{\infty} r^{l}f_n(q_0-r^2)\mathrm{d}r .
\label{eq:moments}
\end{equation}

The thermodynamic parameters of the reservoir involved in thermoelectric transport are derivatives of Eq.~\eqref{eq:moments}. Using the relation
\begin{equation}
	\frac{\mathrm{d}^a N_l (q_0)}{\mathrm{d}q_0 ^a} = N_{l-2a}(q_0)\prod_{i=0}^{a}\left(\frac{l-1}{2}-i \right),\quad a > 0
\label{eq:derivNMoments}
\end{equation}
one finds for the compressibility
\begin{equation}
	\kappa = \left.\frac{1}{k_\mathrm{B}T}\frac{\partial N}{\partial q}\right|_T = \frac{2}{\sqrt{\pi}}\frac{(kT)^2}{(\hbar\bar{\omega})^3}N_0(q_0),
\label{eq:kappa}
\end{equation}
for the dilatation coefficient
\begin{equation}
	\alpha = \frac{1}{\kappa}\left.\frac{\partial N}{\partial T}\right|_{\mu} = k_B\left(6\frac{N_2(q_0)}{N_0(q_0)}-q_0\right)
\label{eq:}
\end{equation}
and for the specific heat
\begin{align}
	C_N = & T\left( \left.\frac{\partial S}{\partial T}\right|_{\mu} - \kappa\alpha^2\right)  \nonumber\\
			= & \frac{8}{\sqrt{\pi}}\left(\frac{k_B T}{\hbar\omega}\right)^3 k_B\left(4N_4(q_0) - 9\frac{N_2^2(q_0)}{N_0}\right).
\label{eq:Cn}
\end{align}

\section{Linear model}

The phenomenological model presented in the text assumes that the particle $I_N = -1/2 \cdot {\rm d}\Delta N/{\rm d}t$ and entropy currents $I_S = -1/2 \cdot {\rm d}\Delta S/{\rm d}t$ are linear functions of the chemical potential and temperature differences \(\Delta \mu\) and \(\Delta T\) between the reservoirs. The symmetry of the transport matrix (Eq.~1) is a consequence of the Onsager reciprocal relations, valid for our system which is microscopically reversible \cite{balian_microphysics_2006_sup}.

The chemical potential and entropy can furthermore be linearized around the equilibrium particle number and temperature by using the compressibility $\kappa = \frac{\partial N}{\partial \mu} \Bigr|_T$, dilatation coefficient $\alpha_\text{r} = \frac{\partial S}{\partial N} \Bigr|_T$ and specific heat $C_N = T \frac{\partial S}{\partial T} \Bigr|_N$ of each reservoir. This allows us to obtain a closed system of first order differential equations on $\Delta N$ and $\Delta T$ \cite{grenier_thermoelectric_2016_sup}:
\begin{equation}
    \tau_N
    \frac{{\rm d}}{{\rm d}t}
    \begin{pmatrix}
    \Delta N  \\
    \Delta T
    \end{pmatrix}
    = - \underline{\Lambda}
    \begin{pmatrix}
    \Delta N \\
    \Delta T 
    \end{pmatrix}
    \text{, }\quad
    \underline{\Lambda} = 
    \begin{pmatrix}
    1 & \alpha \kappa \\
    \frac{\alpha}{l \kappa} & \frac{L + {\alpha}^2}{l}
    \end{pmatrix}
    \label{eq:2by2_NT}
\end{equation}
where $\tau_N = \kappa/2 G$, $\alpha = \alpha_\text{c} - \alpha_\text{r}$ is an effective Seebeck coefficient and $l = C_N/T \kappa$ is the thermodynamic equivalent of a Lorenz number for the reservoirs.

This form of the transport matrix highlights that the particle and temperature dynamics result from a competition between the transport properties of the constriction and the thermodynamical properties of the reservoirs. When $\Delta N = 0$, the particle current can be rewritten as $I_N = G \alpha \Delta T$ and its direction is determined by the sign of $\alpha$: from hot to cold if $\alpha_\text{c} > \alpha_\text{r}$ and from cold to hot if $\alpha_\text{c} < \alpha_\text{r}$.

In absence of thermoelectric coupling $\alpha = 0$, the matrix $\underline{\Lambda}$ is diagonal and the particle and temperature evolutions are decoupled, with timescales $\tau_N$ and $\tau_T = C_N/2 G_T$ respectively. The ratio $L/l = \tau_N/\tau_T$ between the Lorenz numbers of the channel and the reservoirs then indicates which evolution is faster: particle number if $L < l$, temperature if $L > l$. For a non-interacting Fermi gas in the low temperature limit, $l$ approaches the universal value $\pi^2 / 3\cdot k_{\rm B}^2$, and the Wiedemann-Franz law for the channel Lorenz number $L$ implies that the timescales $\tau_N$ and $\tau_T$ should be equal.

In the general case, the time evolutions $\Delta N(t)$ and $\Delta T(t)$ are the sum of two exponential functions,
\begin{widetext}
\begin{gather}
\begin{align}
\Delta N(t) &= \left\{ \frac{1}{2} \Big[e^{-t/\tau_-} + e^{-t/\tau_+} \Big] - \left[1 - \frac{L + \alpha^2}{l} \right] \frac{e^{-t/\tau_-} - e^{-t/\tau_+}}{2(\lambda_+-\lambda_-)} \right\} \Delta N_0 + \frac{\alpha \kappa}{\lambda_+-\lambda_-} \Big[ e^{-t/\tau_-} - e^{-t/\tau_+} \Big] \Delta T_0 \\
\Delta T(t) &= \left\{ \frac{1}{2} \Big[e^{-t/\tau_-} + e^{-t/\tau_+} \Big] - \left[\frac{L + \alpha^2}{l} - 1 \right] \frac{e^{-t/\tau_-} - e^{-t/\tau_+}}{2(\lambda_+-\lambda_-)} \right\} \Delta T_0 + \frac{\alpha}{l \kappa (\lambda_+-\lambda_-)} \Big[ e^{-t/\tau_-} - e^{-t/\tau_+} \Big] \Delta N_0
\end{align}
\label{eq:time_evolution}
\end{gather}
\end{widetext}
with timescales $\tau_\pm = \tau_N/\lambda_\pm$ that depend on the eigenvalues \( \lambda_{\pm} \)of the dimensionless transport matrix $\underline{\Lambda}$:
\begin{equation}
\lambda_\pm = \frac{1}{2} \left(1 + \frac{L+\alpha^2}{l} \right) \pm \sqrt{\frac{\alpha^2}{l} + \left(\frac{1}{2} - \frac{L + \alpha^2}{2 l} \right)^2}.
\label{eq:eigenvalues}
\end{equation}

The ratio between the fast and the slow timescales, plotted in Fig. 3B, can be expanded in the limit $\alpha^2/l \ll 1$ and for $L < l$ as:
\begin{equation}
\frac{\tau_+}{\tau_-} = \frac{L}{l}\left[1-\frac{2 \alpha^2}{l-L} \right].
\label{eq:timescale_ratio}
\end{equation}
The quantity $\frac{\tau_+}{\tau_-} \cdot l$ therefore provides a lower bound for the channel Lorenz number $L$.

\section{Transport coefficients in a non-interacting system}
\begin{figure}
    \center
    \includegraphics{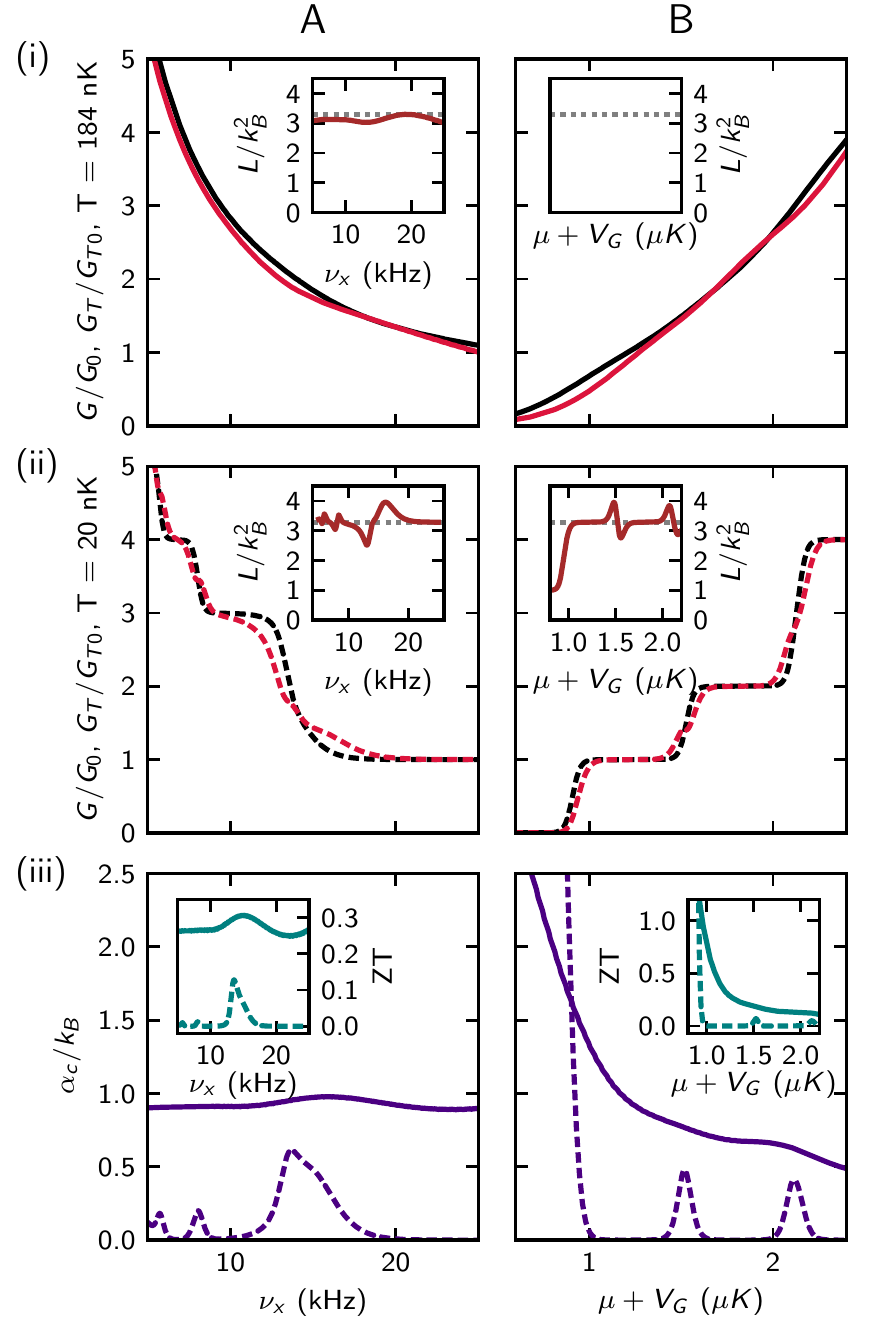}
    \caption{
    Transport coefficients and related ratios for a non-interacting QPC with vertical confinement frequency $\nu_z = \SI{12.9}{\kilo\hertz}$ and reservoir mean chemical potential $\mu = \SI{272}{\nano\kelvin}\cdot k_\mathrm{B}$, \textsl{(A)} as a function of horizontal confinement frequency $\nu_x$ at fixed gate potential $V_G = \SI{1.00(3)}{\nano\kelvin}\cdot k_\mathrm{B}$ and \textsl{(B)} as a function of $\mu + V_G$ at fixed $\nu_x = \SI{20.2}{\kilo\hertz}$.
    (i) Conductance $G$ (black line), thermal conductance $G_T$ (red line) and Lorenz number $L$ (inset) for temperature $T = \SI{184}{\nano\kelvin}$. $G$ and $G_T$ are normalized by the quanta $G_0 = 1/h$ and $G_{T 0} = \pi^2/3 \cdot k_B^2 T /h$ respectively. The Wiedemann-Franz law is indicated by a dotted line at $L/k_B^2 = \pi^3/3$ in the inset. (ii) Id. at temperature $T = \SI{20}{\nano\kelvin}$. (iii) Seebeck coefficient $\alpha_\text{c}$ and figure of merit $ZT = \alpha_\text{c}^2 / L$ (inset) at temperatures $T = \SI{184}{\nano\kelvin}$ (full line) and $\SI{20}{\nano\kelvin}$ (dashed line).
    \label{fig:nonint_theory}
    }
\end{figure}

The theory curves of the Lorenz number and Seebeck coefficient  shown in Figs.~3C and 4C of the main text are based on Landauer theory for a non-interacting Fermi gas. The linear transport coefficients of the constriction $G$, $\alpha_\text{c}$ and $G_T$ at non-zero temperature $T$ are given by the following energy integrals \cite{grenier_thermoelectric_2016_sup}:

\begin{equation}
\label{eq:landauer_g}
G=\frac{1}{h}\int_{-\infty}^{+\infty}d\epsilon\Phi(\epsilon) \left(-\frac{\partial f}{\partial \epsilon} \right)
\end{equation}

\begin{equation}
\label{eq:landauer_alpha}
G\alpha_{\text{c}}=\frac{1}{hT}\int_{-\infty}^{+\infty}d\epsilon(\epsilon-\mu)\Phi(\epsilon) \left(-\frac{\partial f}{\partial \epsilon} \right)
\end{equation}

\begin{equation}
\label{eq:landauer_gt}
\frac{G_{T}}{T}+G\alpha_{\text{c}}^{2}=\frac{1}{hT^{2}}\int_{-\infty}^{+\infty}d\epsilon(\epsilon-\mu)^{2}\Phi(\epsilon) \left(-\frac{\partial f}{\partial \epsilon} \right),
\end{equation}

where $\mu$ is the mean chemical potential of the reservoirs, $f(\epsilon) = 1/[1 + \exp((\epsilon - \mu)/k_\text{B} T)]$ is the Fermi-Dirac distribution function and $\Phi(\epsilon)$ is the energy-dependent transmission function of the constriction, indicating how many modes are available for transport at energy $\epsilon$. For a ballistic quantum point contact,
\begin{equation}
\Phi(\epsilon) = \sum_{n, m} \Theta (\epsilon - \epsilon_{n, m})
\end{equation}
where $\epsilon_{n, m} = (1/2 + n) h \nu_x + (1/2 + m) h \nu_z - V_G$ is the energy of the transverse mode labelled by integers $n$ and $m$ in the $x$- and $z$-directions, and lowered by the attractive gate potential $V_G$.
The integrals (\ref{eq:landauer_g}), (\ref{eq:landauer_alpha}) and (\ref{eq:landauer_gt}) can be related to the individual transport coefficients $G$, $\alpha_\text{c}$ and $G_T$, shown in Fig. \ref{fig:nonint_theory} as a function of transverse confinement frequency $\nu_x$ and local chemical potential $\mu + V_G$ for the QPC parameters used in Figs. 3 and 4 in the main text.

At the temperature $T = \SI{184}{\nano\kelvin}$ realized experimentally [subpanels (i)], both conductance $G$ and thermal conductance $G_T$ show a smooth evolution with respect to $\nu_x$ and $\mu + V_G$ and display an almost constant ratio that results in a Lorenz number $L = G_T/T G$ close to the Wiedemann-Franz (WF) law value \(L_{\rm WF}=\pi^2 / 3\cdot k_{\rm B}^2\). A departure from this value can be observed when the degeneracy in the QPC is reduced (or equivalently, for low conductance $G$) at chemical potentials $\mu + V_G < \SI{1.3}{\micro\kelvin}\cdot k_\mathrm{B}$. Decreasing temperature by one order of magnitude to $T = \SI{20}{\nano\kelvin}$ [subpanels (ii)] is necessary to resolve the quantization of the conductance in units of $G_0 = 1/h$ and of the thermal conductance in units of $G_{T0} = \pi^3/3 \cdot k_B^2 T /h$. There the Lorenz number shows additional oscillations around the WF value in-between the conductance plateaus \cite{houten_thermo-electric_1992_sup}. The conductance jumps go along with peaks in the Seebeck coefficient $\alpha_\text{c}$ [subpanels (iii), purple curves] at low temperatures $T = \SI{20}{\nano\kelvin}$ (dashed lines), which are smoothened out at large $T = \SI{184}{\nano\kelvin}$ (full lines). Since the Lorenz number $L$ barely varies over both parameter ranges, the variations of the thermoelectric figure of merit $ZT = \alpha_\text{c}^2 / L$ [subpanels (iii), inset] are mostly determined by those of $\alpha_\text{c}$. Except for low conductance regimes, $ZT$ is expected to remain below one and hence contrasts with the larger value obtained experimentally at unitarity, which is associated to an anomalously low Lorenz number.

\section{Efficiency}

In the first part of its evolution, mechanical work is released by the reservoirs, $P = I_N \Delta \mu > 0$, while thermal power is extracted from the cold reservoir, $Q_c = - T_c I_S > 0$. Our system can therefore be viewed as a thermoelectric cooler, associated with a maximal coefficient of performance \cite{goupil_thermodynamics_2011_sup}
\begin{equation}
	\varphi_\text{max} = \frac{Q_c}{P} = \varphi_\text{C} \frac{\sqrt{1+ZT} - T_\text{h}/T_\text{c}}{\sqrt{1+ZT} + 1},
\end{equation}
which is related to the thermoelectric figure-of-merit $ZT = \alpha_\text{c}^2/L$ and the Carnot factor
\begin{equation}
	\varphi_\text{C} = \frac{T_\text{c}}{T_\text{h} - T_\text{c}} = 2.4.
\end{equation}
The estimates for the Seebeck coefficient $\alpha_\text{c}$ and Lorenz number $L$ for Fig. 2 and 3 yield $\varphi_\text{max}$ of 1.2, i.e. $51 \%$ of the Carnot bound $\varphi_\text{C}$.

When the particle current reverts, the heat flow from hot to cold $Q = - \Delta I_S \Delta T < 0$ is converted into work, $P < 0$. This heat engine regime is associated with the maximum efficiency
\begin{equation}
\eta_\text{max} = \frac{P}{Q} = \eta_\text{C} \frac{\sqrt{1+ZT} - 1}{\sqrt{1+ZT} + T_\text{c}/T_\text{h}}
\end{equation}
with the Carnot efficiency
\begin{equation}
	\eta_\text{C} = \frac{T_\text{h} - T_\text{c}}{T_\text{h}} = 0.29.
\end{equation}
This amounts to an absolute value of 0.18 using the previous estimates, i.e. $63 \%$ of the Carnot bound $\eta_\text{C}$.

\section{Evaluation of transport parameters}
To describe the relaxation dynamics of the temperature-biased system, we rely on finding the transport coefficients that characterize the transport matrix \cite{balian_microphysics_2006_sup}
\begin{equation}
    \begin{pmatrix}
    I_N  \\
    I_S  
    \end{pmatrix}
    = 
    G
    \begin{pmatrix}
    1                   &   \alpha_{\rm{ch}} \\
    \alpha_{\rm{ch}}    & L + \alpha_{\rm{ch}}^2
    \end{pmatrix}
    \cdot
    \begin{pmatrix}
    \Delta\mu\\
    \Delta T 
    \end{pmatrix},
    \label{eq:2by2_SI}
\end{equation}
and the reservoir parameters appearing in Eq.~\eqref{eq:2by2_NT}.
The thermodynamic coefficients \( \kappa,\alpha_\mathrm{r},C_N \) of the reservoirs are known from the equation of state.
In the following we describe how we extract the transport coefficients of the channel \( \alpha_\mathrm{c},\ G \) and \( G_T \) from the time evolution of \(\Delta N\) and \(\Delta T\).

\subsection{Seebeck coefficient}

The Seebeck coefficient is defined as the ratio
\begin{equation}
	\alpha_{{\rm c}}=-\left.\frac{\Delta\mu}{\Delta T}\right|_{I_{N}=0}.
	\label{eq:seebeck}
\end{equation}
For a given confining frequency of the QPC $\nu_{x}$ and value of
the gate potential $V_{{\rm G}}$, the evolution can either exhibit
a non-equilibrium steady state or show a significant amount of relaxation
towards the final equilibrium state (see main text Fig.~3A). Thus, for
each value of the gate potential in Fig.~4C, we measure $\Delta N(t)$
to determine, using a biexponential fit, the time $t_{{\rm stop}}(V_{{\rm G}})$
at which the particle current goes to zero before reversing sign. In the curves shown in Fig.~4a of the main text, this point corresponds to the maximum in particle difference \(\Delta N\).

We fix the constriction parameter $\nu_{x}$ and $V_{{\rm G}}$
as well as the transport time $t_{{\rm stop}}(V_{{\rm G}})$. Then we
record the state of the the gas ($\Delta N$, $\Delta T$, $\Delta\mu$)
after $t_{{\rm stop}}(V_{{\rm G}})$ for different modulation amplitudes
of the heating beam which corresponds to different initial temperature
differences. In the following paragraph, the different preparation conditions are indicated with an index $i$. Each measurement is repeated at least five times.


For all measurements performed with the preparation condition $i$ we obtain a distribution of points in the \(\Delta N, \Delta T\)-
plane which we can approximate by a mean value with independent Gaussian errors on $\Delta N$
and $\Delta T$. We use the equation of state of the unitary Fermi
gas to convert these data to the $(\Delta T$, $\Delta\mu)$-plane,
where the validity of Eq.~\eqref{eq:seebeck} can be checked. There,
the spread of the measurements stemming from the same preparation
procedure still follows a Gaussian distribution with mean values $\overline{\Delta\mu}_{i}$ and $\overline{\Delta T}_{i}$. The spread of the measurements is again described by a bivariate Gaussian distribution with covariance matrix
\[
\Sigma_{i}=\left(\begin{array}{cc}
\sigma_{\mu\mu}^{i} & \sigma_{\mu T}^{i}\\
\sigma_{\mu T}^{i} & \sigma_{TT}^{i}
\end{array}\right)
\]
where
$\left\langle \left(\Delta\mu\right)^{2}\right\rangle -\left\langle \Delta\mu\right\rangle ^{2}=\sigma_{\mu\mu}^{i}$,
$\left\langle \left(\Delta T\right)^{2}\right\rangle -\left\langle \Delta T\right\rangle ^{2}=\sigma_{TT}^{i}$
and $\left\langle \Delta\mu\Delta T\right\rangle -\left\langle \Delta\mu\right\rangle \left\langle \Delta T\right\rangle =\sigma_{\mu T}^{i}$.
The off-diagonal elements of $\Sigma_{i}$ are not constrained to zero, and the orientation and spread of the Gaussian
distribution depends on the preparation condition (see Fig.~4B).

It is therefore necessary to adapt the ordinary least square algorithm
to take into account this unusual error distribution; for instance,
neglecting the error along the \(x\)-axis can lead to
an underestimation of the parameters \cite{fuller_measurement_2009_sup}.


For a given data point (i.e. a given preparation condition) $M_{i}=\left(\overline{\Delta T}_{i},\overline{\Delta\mu}_{i}\right)$,
we redefine its distance to any point $P=(x,y)$ by
\begin{equation}
\left|M_{i}P\right|_{\sigma^{i}}^{2}=\left(\overline{\Delta T}_{i}-x,\overline{\Delta\mu}_{i}-y\right)\Sigma_{i}^{-1}\left(\begin{array}{c}
\overline{\Delta T}_{i}-x\\
\overline{\Delta\mu}_{i}-y
\end{array}\right)
\end{equation}
which takes into account the uncertainty on the data, and where
\begin{equation}
\Sigma_{i}^{-1}=\frac{1}{\sigma_{\mu\mu}^{i}\sigma_{TT}^{i}-\left(\sigma_{\mu T}^{i}\right)^{2}}\left(\begin{array}{cc}
\sigma_{\mu\mu}^{i} & -\sigma_{\mu T}^{i}\\
-\sigma_{\mu T}^{i} & \sigma_{TT}^{i}
\end{array}\right).
\end{equation}
That way, we can compute the distance of $M_{i}$ to a given slope with intercept $\beta_0$ and slope $\beta_1$ by minimizing $\left|M_{i}P\right|_{\sigma^{i}}^{2}$ with $P=(x,\beta_0+\beta_1 x)$ over \( x \), which yields
$$ \min_x(\left|M_{i}P\right|_{\sigma^{i}}^{2}) = \frac{\left(\overline{\Delta\mu}_{i}-\beta_{0}-\beta_{1}\overline{\Delta T}_{i}\right)^{2}}{\sigma_{\mu\mu}^{i}-2\beta_{1}\sigma_{\mu T}^{i}+\beta_{1}^{2}\sigma_{TT}^{i}}$$

The fitting procedure then consists in numerically minimizing the sum \( S \) of these distances over $\beta_0$ and $\beta_1$ 
\begin{equation}
S=\sum_{i}\left|M_{i}P_{i}\right|_{\sigma^{i}}^{2}=\sum_{i}\frac{\left(\overline{\Delta\mu}_{i}-\beta_{0}-\beta_{1}\overline{\Delta T}_{i}\right)^{2}}{\sigma_{\mu\mu}^{i}-2\beta_{1}\sigma_{\mu T}^{i}+\beta_{1}^{2}\sigma_{TT}^{i}}
\end{equation}
The slope can then be identified with the opposite of the Seebeck coefficient.

The error on the coefficient is obtained by a bootstrap analysis. Each point is replaced by a randomly drawn point with a normal bivariate distribution with mean $(\overline{\Delta\mu}_{i},\overline{\Delta T}_{i})$ and covariance ${\Sigma}^{i} $ and a new slope $\beta_1^j$ is computed. This is repeated 3000 times, and yields a distribution which can be fitted by a Gaussian distribution whose width is the error on the slope. We have checked that this method is consistent with the analytical formula in the case of identical covariance matrices for all points.

\subsection{Particle conductance}
\begin{figure*}[htb]
	\centering
		\includegraphics{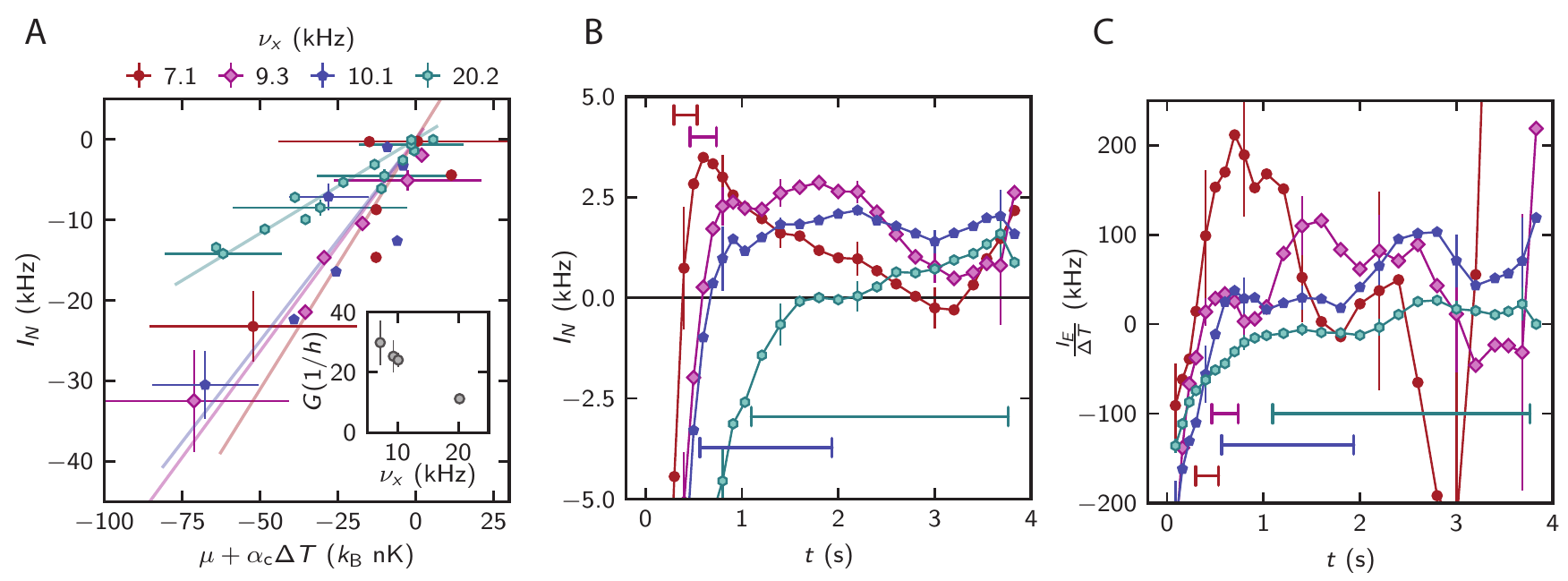}
	\caption{Particle and energy currents for different \(\nu_x\). 
	\textsl{(A)} Current-bias characteristics for several values of the horizontal wire frequency \(\nu_x\). The solid lines are linear fits through the origin, whose slopes are equal to the conductance \(G\) (see Eq.~\eqref{eq:G_def}). Points with \(I_N > 0 \) and \(\Delta \mu + \alpha_\mathrm{c}\Delta T>0\) are suppressed from the linear fit, as their distribution is dominated by noise in \(\Delta \mu\).
	\textsl{(B)} Particle current \( I_N \) versus time \(t\). 
	\textsl{(C)} \(I_E / \Delta T\) versus time \(t\). The bars in \textsl{(B)} and \textsl{(C)} indicate for each value of confinement frequency the time intervals where the particle current is less than two standard deviations away from zero.
	}
	\label{figS:IV_IN_IE}
\end{figure*}

Following Eq.~\eqref{eq:2by2_SI}, the particle conductance used to estimate the Lorenz number in Fig.~3 of the main text is
\begin{equation}
	G=\frac{I_N}{\Delta \mu + \alpha_\mathrm{c}\Delta T}.
\label{eq:G_def}
\end{equation}
The Seebeck coefficient \(\alpha_\text{c}\) as defined by Eq.~\eqref{eq:seebeck} is obtained from the points of vanishing particle current.
The particle current \(I_N = -1/2 \cdot {\rm d}\Delta N/\mathrm{d}t\) is obtained by evaluating the instantaneous time derivative of the particle number difference \(\Delta N\). To this end we apply a sliding window derivation method with a Gaussian weighting function of standard deviation \(\sigma=0.3\)~s, truncated after 3 points away from the center point. The resulting time-dependent current \(I_N\) plotted versus \( \Delta \mu + \alpha_\mathrm{c}\Delta T\) is shown as current-bias characteristics in Fig.~\ref{figS:IV_IN_IE}A. A linear fit to the data confirms the validity of Eq.~\eqref{eq:G_def}, and the slope yields directly the conductance \( G \) shown as an inset in Fig.~\ref{figS:IV_IN_IE}.

\subsection{Thermal conductance}
From the analysis of the equation of state, we extract the energy of each reservoir, the energy difference \( \Delta E \) and the energy current \(I_E = -1/2\cdot \mathrm{d}\Delta N / \mathrm{d}t\). According to Eq.~(2) in the main text, \(G_T\) is the ratio between energy current \(I_E\) and temperature difference \(\Delta T\) at the time where \(I_N=0\).
This time is known up to the accuracy in measuring \(I_N\). The uncertainty in \(I_N\) is limited by the error originating from the numerical derivative evaluation, which --- averaged over all four confinement frequencies \(\nu_x\) and transport times \(t\) --- is around \(1.0\)~kHz. Consequently for each value of confinement \(\nu_x\), we obtain an interval in transport time \(t\) over which we average \(I_E/\Delta T\) to determine \(G_T\). The averaging is weighted with a Gaussian function of standard deviation \(\sigma\) given by half the time interval shown in Fig.~\ref{figS:IV_IN_IE}, and centered on the point of vanishing current obtained from the fits shown in Fig.~3A of the main text.

\section{Density distribution in the center}

As a result of our geometry, high density regions form at the entrance and exit of the QPC and change the nature of the gas from normal to superfluid. We refer to these regions as particle pockets and characterize them in the following.

\subsection{Effective potential}
Since transport is directed along a single direction in our structure it is convenient to separate longitudinal (\(y\)) and transverse (\(x, z\)) coordinates. The motion in the transverse directions is prohibited by strong harmonic confinements that lead to quantized states with the following eigenenergies:
\begin{align}
    E_{n_x} (y) = h \nu_x f_x (y) (n_x + 1/2) \\
    E_{n_z} (y) = h \nu_z f_z (y) (n_z + 1/2)
\end{align}
The envelope functions \(f_x\) and \(f_z\) account for the varying transverse trapping frequencies due to the Gaussian envelopes of the beams and are summarized in Table~\ref{tab:envelope-functions}. 

The anharmonicity of the transverse confinements are small in the range where we rely on the harmonic description. It is largest in the pockets away from the center and for higher transverse modes where particles explore a wider region of the trapping potentials. To quantify this effect we numerically calculate the eigenenergies for the exact potentials along the \( x \) and \( z \) directions separately and find that the harmonic energies are only \SI{1}{\percent} larger than the exact ones for the state \( n_x = 15 \) and \SI{3}{\percent} for \( n_z = 15 \).

As long as the variation of the transverse frequencies between channel and reservoirs is adiabatic, the particles remain in the same transverse state during their motion. Hence, for the longitudinal motion the transverse energy acts as an additional contribution to the potential \(V (y)\) along the transport direction, giving rise to the effective potential \( V_\text{eff} (y) = E_{n_x} (y) + E_{n_z} (y) + V(y) \). In our setup the potential \(V (y)\) consists of the attractive gate and the magnetic trap:
\begin{align}
    V_\text{G} (y) = - V_\text{G} f_g(y) \\
    V_\text{M} (y) = \frac{1}{2} m \omega_y^2 y^2 .
\end{align}

\begin{table}
    \begin{tabular}{lll}
        Envelope function & Waist & Description \\
        \hline
        \(f_x(y) = \exp(-y^2 / w_x^2)\) & \(w_x = \SI{5.5}{\micro\meter}\) & \(x\) confinement \\
        \(f_z(y) = \exp(-y^2 / w_z^2)\) & \(w_z = \SI{30.2}{\micro\meter}\) & \(z\) confinement \\
        \(f_g(y) = \exp(-2 y^2 / w_g^2)\) & \(w_g = \SI{34.5}{\micro\meter}\) & gate potential \\
        \hline
    \end{tabular}
    \caption{Envelope functions determining the effective one-dimensional potentials}
    \label{tab:envelope-functions}
\end{table}

\begin{figure}
    \includegraphics{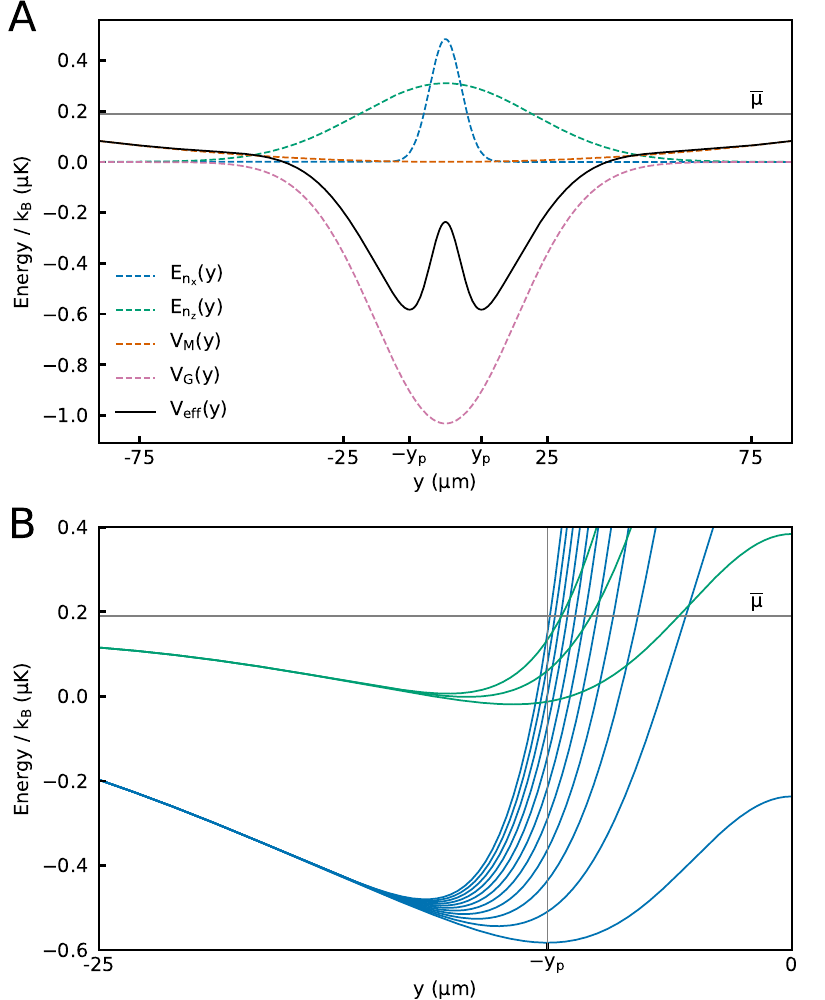}
    \caption{Effective one-dimensional potential. \textsl{(A)} Its contributions are the transverse energies \( E_{n_x} \) and \( E_{n_z} \) plotted in the ground state \( n_x = 0 \) and \( n_z = 0 \) and with trapping frequencies \( \nu_x = \SI{20.2}{\kilo\hertz} \) and \( \nu_z = \SI{12.9}{\kilo\hertz} \), the harmonic trap \( V_M \) with frequency \( \omega_y = 2\pi \cdot \SI{28.3}{\hertz} \) and the Gaussian gate potential \( V_\text{G} \) with amplitude \( V_G = \SI{1.0}{\micro\kelvin} \cdot k_B \). \textsl{(B)} Effective potential for different transverse modes \( (n_x, n_z) \) that are occupied in the particle pockets. The blue and green curves distinguish modes with \( n_z = 0 \) and \( n_z = 1 \). They fan out close to the QPC due to different numbers \( n_x \). The indicated chemical potential \( \bar{\mu} \) imposed by the reservoirs is \( \SI{190}{\nano\kelvin} \cdot k_B \).}
    \label{fig:effective-potential}
\end{figure}

Figure \ref{fig:effective-potential}A shows the effective potential and its contributions in the transverse ground state \( n_x = 0 \) and \( n_z = 0 \) for the parameters in Fig.~2. The particle pockets at positions \( \pm y_p \) are clearly visible in the vicinity of the constriction.

\subsection{Superfluid transition}

To extract the superfluid transition temperature in the pockets as indicated in Fig.~4C we locally apply the thermodynamics of a homogeneous unitary Fermi gas. Namely, from the local Fermi temperature \( \breve{T}_F =  \hbar^2 / (2 m) (6 \pi^2 \breve{n})^{2/3} / k_B \) we obtain the critical temperature \( \breve{T}_c \) of \( 0.167 \cdot \breve{T}_F \) \cite{ku_revealing_2012_sup}. The density~\( \breve{n} \) in the pockets is given by the known equation of state \eqref{eq:EoSn}, the average temperature \( \bar{T} \) and the local chemical potential \( \breve{\mu} = \bar{\mu} - V_\text{eff} (y_p) \) which is calculated from the effective potential. The pockets are superfluid for all measurements shown in the paper, except for the points of strongest heating used to determine the first two data points in Fig.~4C.

By applying the equation of state Eq.~\eqref{eq:EoSn} to the pockets, we implicitly assume that they are three-dimensional and the gas remains thermalized during the transport process. In the following we justify these assumptions.

\subsection{Dimensionality}

For noninteracting particles the dimensionality of the pockets can be inferred from the occupied transverse modes, i.e.\ modes \( (n_x, n_z) \) with an effective potential \( V_\text{eff} (y_p) \) smaller than the chemical potential \( \bar{\mu} \), as shown in Fig.~\ref{fig:effective-potential}B. Far away from the QPC the transverse confinements are weak and the modes are almost degenerate. Towards the center they first split due to the confinement along \( z \) and then fan out as the trapping along \( x \) increases. The plot indicates that at most 2 modes are occupied along the \( z \)-direction, and 11 modes along \( x \) for typical experimental parameters. Hence, the pockets are quasi-two-dimensional. With interactions particles tend to populate higher excited modes, bringing the gas closer to three-dimensions \cite{Dyke2016_sup}.

\subsection{Thermalization}

For the particle pockets to be in thermal equilibrium, it is necessary that the three-dimensional regions of the reservoirs are equilibrated as they are in close contact. To check this, we compare the overall transport time with the interparticle collision time in the reservoirs, which can be determined through a Boltzmann approach in a harmonically trapped unitary Fermi gas as \cite{Gehm2003_0_sup}
\begin{equation}
    \tau = \tau_0 / I (T/T_F)
\end{equation}
with the dimensionless collision integral \( I (T/T_F) \) and the natural time \( \tau_0 = 6 \pi \hbar / E_F \). The Fermi energy in a harmonic trap reads \( E_F = k_B T_F = h \bar{\nu}_\mathrm{r} (6 N)^{1/3} \). For typical atom numbers \( N = 97 \cdot 10^3 \) and temperatures \( T = \SI{210}{\nano\kelvin} \) the degeneracy \( T / T_F \) is 0.4 which translates into a dimensionless collision integral \( I (T/T_F) \) of 4.6. Together with the natural time \( \tau_0 \) of \SI{295}{\micro\second} this results in a collision time \( \tau \) of \SI{64}{\micro\second}, which is short compared to the overall transport time of \SI{4}{\second} and the time scales \( \tau_+ \) and \( \tau_- \).

\section{Transport modes}

The effective potential allows us to separate transport into contributions of independent transverse modes in the channel as long as transport happens adiabatically (see section ``Effective potential''). A mode contributes if the corresponding effective potential is below the chemical potential \( \bar{\mu} \).

In the situation of Fig.~2 and 3 for the strongest confinement only the lowest transverse mode is available. While decreasing the confinement up to four modes become occupied. In Fig.~4 the attractive gate lowers the effective potential and the available modes vary from one up to four.

\end{document}